%% file: main.tex
\documentclass[letterpaper, 10 pt, conference]{ieeeconf}  % Comment this line out
\IEEEoverridecommandlockouts                              % This command is only
\usepackage{graphics} % for pdf, bitmapped graphics files
\usepackage{epsfig} % for postscript graphics files
\usepackage{mathptmx} % assumes new font selection scheme installed
\usepackage{times} % assumes new font selection scheme installed
\usepackage{amsmath} % assumes amsmath package installed
\usepackage{amssymb}  % assumes amsmath package installed
\usepackage{algorithmicx}
\usepackage{xcolor}
\usepackage{xspace}
\usepackage{bm}
\usepackage{cite}
\usepackage{algorithm,algpseudocode,setspace}
\usepackage{soul}
\usepackage{pgfplots}
\usepgfplotslibrary{groupplots}

\newtheorem{remark}{Remark}
\newtheorem{theorem}{Theorem}

\newtheorem{lemma}{Lemma}
\newtheorem{definition}{Definition}

\DeclareMathOperator{\tr}{Tr}

\usepackage{todonotes}

\title{\LARGE \bf
A Modified Adaptive Data-Enabled Policy Optimization Control to Resolve State Perturbations
}

\author{Mojtaba Kaheni, Niklas Persson, Vittorio De Iuliis, Costanzo Manes, and Alessandro V. Papadopoulos% <-this % stops a space
\thanks{This work was supported in part by the Swedish Research Council (VR) with grant ``Pervasive Self-Optimizing Computing Infrastructures (PSI)'' n. 2020-05094, and by the Knowledge Foundation (KKS) with grant ``Mälardalen University Automation Research Center (MARC)'', n. 20240011. The work was also supported in part by the Italian Government under CIPE resolution n. 70/2017 (Centre of excellence EX-EMERGE).}
\thanks{M. Kaheni, N. Persson, and A.V. Papadopoulos are with the Division of Intelligent Future Technologies, M\"alardalen University, 721 23 V\"aster{\aa}s, Sweden.    (e-mails: {\tt\footnotesize mojtaba.kaheni@mdu.se, niklas.persson@mdu.se, alessandro.papadopoulos@mdu.se}).%
}
\thanks{V. De Iuliis and C. Manes are with the Department of Information Engineering, Computer Science and Mathematics, University of L'Aquila, Italy. (e-mails: {\tt\footnotesize vittorio.deiuliis@univaq.it, costanzo.manes@univaq.it}).}}

\begin{document}

\maketitle
\thispagestyle{empty}
\pagestyle{empty}

%%%%%%%%%%%%%%%%%%%%%%%%%%%%%%%%%%%%%%%%%%%%%%%%%%%%%%%%%%%%%%%%%%%%%%%%%%%%%%%%
% \alessandro{In the title, should we mention something on the fact that it is an adaptive controller, similarly to what we did in the TCST paper?
% Mojtaba: Added}
\begin{abstract}
This paper proposes modifications to the data-enabled policy optimization (DeePO) algorithm to mitigate state perturbations. DeePO is an adaptive, data-driven approach designed to iteratively compute a feedback gain equivalent to the certainty-equivalence LQR gain. Like other data-driven approaches based on Willems' fundamental lemma, DeePO requires persistently exciting input signals. However, linear state-feedback gains from LQR designs cannot inherently produce such inputs. To address this, probing noise is conventionally added to the control signal to ensure persistent excitation.
However, the added noise may induce undesirable state perturbations.
We first identify two key issues that jeopardize the desired performance of DeePO when probing noise is not added: the convergence of states to the equilibrium point, and the convergence of the controller to its optimal value.
To address these challenges without relying on probing noise, we propose Perturbation-Free DeePO (PFDeePO) built on two fundamental principles. 
First, the algorithm pauses the control gain updating in DeePO process when system states are near the equilibrium point. 
Second, it applies a multiplicative noise, scaled by a mean value of $1$ as a gain for the control signal, when the controller converges. This approach minimizes the impact of noise as the system approaches equilibrium while preserving stability.
We demonstrate the effectiveness of PFDeePO through simulations, showcasing its ability to eliminate state perturbations while maintaining system performance and stability.
\end{abstract}

%%%%%%%%%%%%%%%%%%%%%%%%%%%%%%%%%%%%%%%%%%%%%%%%%%%%%%%%%%%%%%%%%%%%%%%%%%%%%%%%
\section{Introduction}

Traditionally, the design of controllers has relied on two primary approaches: first-principle modeling which is based on established physical laws and principles specific to the system's domain,
% \alessandro{first-principle models?
% Mojtaba: added. }
or system identification techniques that use available data to construct a mathematical representation. These mathematical models, often dynamic in nature, are then employed to analyze the system's behavior under various inputs and to derive a control law that satisfies specified performance criteria.

However, the seminal work by Willems \textit{et al.}~\cite{WILLEMS2005} introduced a groundbreaking concept with far-reaching implications for control system design. This study revealed that a finite set of system trajectories, generated using persistently exciting inputs, is sufficient to describe the complete behavior of a controllable linear time-invariant (LTI) system. It implies that controllable LTI systems can be fully characterized using only finite historical data, eliminating the need for traditional model-based representations.
This result has sparked widespread interest within the control systems community, as it presents a promising alternative for simplifying controller design. By potentially circumventing the expensive and time-consuming processes of system identification or first-principle modeling, this approach offers a new framework for streamlining and accelerating the development of control systems.

Linear quadratic regulators (LQR) are a widely adopted control design method due to their ability to balance the trade-off between convergence rate and control effort, which can be adjusted based on the designer's preferences. In LTI systems, when the mathematical state-space representation of the system is available, solving a straightforward optimization problem provides the state-feedback gain that minimizes the corresponding cost function \cite{chen2004linear}. In cases where, instead of the state-space representation, sufficiently rich measurements are available, it is still possible to design an LQR controller by conventional approaches, but an initial identification step is required to first find the system matrices. This approach is typically referred to in the literature as {\it{indirect data-driven}} LQR design \cite{Cohen2019, Horia2019,Ferizbegovic2020, Sforni2023}.  

In contrast, several methods have recently been proposed to design LQR controllers for LTI systems directly from data, without the need to identify the system matrices \cite{Tesi2020,Mohammadi2021, Tesi2021, Dorfler2022, Lopez2023, Dorfler2023, Fan2024, Persson2024}. These approaches are commonly known as {\it{direct data-driven}} LQR design. Most of the direct data-driven LQR design methods rely on pre-measured datasets to compute the optimal feedback gain.
A natural consideration is to allow the control system to leverage the data it measures during operation to improve or fine-tune the LQR performance.
% \alessandro{I would reformulate the question to be a sentence. Something like: ``An adaptive strategy can be more effective in leveraging the data collected at runtime to improve and fine-tune the LQR performance.'' Then modify the next paragraph to remove ``to address this question''.
% Mojtaba: Done.}
Toward this goal, the Data-enabled Policy Optimization (DeePO) algorithm was recently proposed \cite{zhao2023data, zhao2024data}. DeePO incorporates an adaptation feature on top of direct data-driven LQR design. At each time step, newly measured input and state data are added to the previously stored dataset, and the control feedback gain is updated iteratively using a learning rate, steering the design towards reducing the objective function's cost based on the updated data.  
Furthermore, it has been proven that DeePO computes a feedback gain equivalent to the certainty-equivalence LQR gain typically derived using indirect data-driven techniques \cite{zhao2024data}. 
DeePO has been evaluated in simulations on a power converter system~\cite{zhao2024direct},
and was used for balancing an autonomous bicycle in real experiments~\cite{persson2025adaptive}. 
Similar to other data-driven methods based on Willems' fundamental lemma, DeePO requires persistently exciting input signals. On the other hand, linear state-feedback gains from LQR designs cannot inherently generate such inputs. To overcome this limitation, probing noise is conventionally added to the control signal to ensure persistent excitation. This added noise often induces undesirable state perturbations. In this article, we introduce PFDeePO to resolve state perturbations in DeePO.

%The remainder of this article is organized as follows: In Section \ref{sec:DeePO}, we briefly revisit the DeePO algorithm and discuss the reasons it leads to state perturbations. Our proposed algorithm to address state perturbations is introduced in Section \ref{sec:Algos}. Simulation results are presented in Section \ref{sec:Simulations}, and Section \ref{sec:Conclusion} concludes the article.
\subsection{Notation}
Throughout this paper, unless clearly stated otherwise, the symbols $\mathbb{N}$, $\mathbb{Z}$, and $\mathbb{R}$ denote the sets of, natural, integer, and real numbers, respectively.
%\costanzo{$\mathbb{N}$ is used for {\it naturals}, i.e.\ positive integers.
%I think that $\mathbb{Z}$ is the standard notation for integers.}
Scalars are represented by lowercase letters such as $x$, while $\mathbf{x}$ and $\mathbf{X}$ denote a (column) vector and a matrix, respectively. The notation $\mathbf{X} \prec 0$ ($\mathbf{X} \preceq 0$) and $\mathbf{X} \succ 0$ ($\mathbf{X} \succeq 0$) indicates that $\mathbf{X}$ is negative (semi-) definite and positive (semi-) definite, respectively. Additionally, $\mathbf{I}_i$ denotes the $i \times i$ identity matrix. The symbols $\mathbf{0}_{i \times j}$ and $\mathbf{1}_{i \times j}$ represent all-zeros and all-ones matrices of size $i \times j$, respectively. The 2-norm of matrix $\mathbf{X}$ is denoted by $\|\mathbf{X}\|$. $\mathbf{X}^{\top}$, $\tr(\mathbf{X})$, and $\mathbf{X}^\dagger$ represent the transpose, trace, and pseudoinverse of matrix $\mathbf{X}$, respectively. The notation $\mathbf{x}^i$ and $\mathbf{X}^i$ refer to the $i$-th coordinate of $\mathbf{x}$ and the $i$-th row of $\mathbf{X}$, respectively. Meanwhile, $\mathbf{X}^{i,j}$ denotes the element located in the $i$-th row and $j$-th column of $\mathbf{X}$. $\mathcal{N}(\mathbf{x}, \mathbf{X})$ represents a multivariate Gaussian distribution with a mean vector $\mathbf{x}$ and a covariance matrix $\mathbf{X}$. $\Pi_{\mathbf{X}}$ stands for projection operator on $\mathbf{X}$. Finally, $\underline{\sigma}(\mathbf{X})$ denotes the smallest singular value of $\mathbf{X}$.
%\vitt{Comments (no need to address before submission): making vectors and matrices bold is unnecessary in my opinion, and makes everything bulkier. More impotant: adding element index as superscript is highly unusual and is prone to contraddiction/confusion with powers: just consider equation (1)... definitely better to use $x(k)$ and $u(k)$ and save subscripts for elements of vector/matrices}

\section{Data-Driven Policy Optimization for LQR Learning}
\label{sec:DeePO}
% \alessandro{Maybe introduce a subsection called Background

% Mojtaba: Done}
\subsection{Background}
Consider an LTI discrete-time system, represented in state space form as:
\begin{align}
    \begin{aligned}
        \mathbf{x}_{k+1} &= \mathbf{A} \mathbf{x}_{k} + \mathbf{B} \mathbf{u}_{k} + \boldsymbol{\omega}_{k}, \\
        \mathbf{z}_{k} &= \begin{bmatrix}
            \mathbf{Q}^{1/2} & \mathbf{0}_{n \times m} \\ \mathbf{0}_{m \times n} & \mathbf{R}^{1/2}
        \end{bmatrix}  \begin{bmatrix}
            \mathbf{x}_{k} \\ \mathbf{u}_{k}
        \end{bmatrix},
    \end{aligned}
    \label{eq:system}
\end{align}
where $k \in \mathbb{N}$ is the index for counting samples, $\mathbf{x} \in \mathbb{R}^n$ is the state, $\mathbf{u} \in \mathbb{R}^m$ represents the input, and $\boldsymbol{\omega}_k$ is noise.
% \alessandro{Since we are using bold symbols for vectors, should we use $\boldsymbol{\omega}_k$ also bold? Note that $\boldsymbol{\omega}_k$ does not make it bold.
% Mojtaba: Done!}
Furthermore, let $\mathbf{z}_{k} \in \mathbb{R}^{n+m}$ represent the performance signal. We assume that the pair $(\mathbf{A}, \mathbf{B})$ is controllable, and that $(\mathbf{Q}, \mathbf{R})$ are positive definite square matrices with compatible dimensions. The objective of the LQR design is to determine a state feedback controller, $\mathbf{K} \in \mathbb{R}^{m \times n}$, that minimizes the $\mathcal{H}_2$-norm of the transfer function $\mathcal{T}(\mathbf{K}) :\boldsymbol{\omega} \mapsto \mathbf{z}$ of the following closed-loop system:
\begin{equation}
\label{eq:LQR1}
    \begin{bmatrix}
        \mathbf{x}_{k+1} \\ \mathbf{z}_{k}
    \end{bmatrix} =
    \left[
    \begin{array}{c|c}
        \mathbf{A} + \mathbf{B}\mathbf{K} & \mathbf{I}_{n} \\ 
        \hline
        \begin{bmatrix}
            \mathbf{Q}^{1/2} \\
            \mathbf{R}^{1/2}\mathbf{K}
        \end{bmatrix} & \mathbf{0}_{(m+n)\times n}
    \end{array}
    \right]
    \begin{bmatrix}
        \mathbf{x}_{k} \\ \boldsymbol{\omega}_{k}
    \end{bmatrix}.
\end{equation}
As discussed in \cite{anderson2007optimal}, the $\mathcal{H}_2$-norm of the transfer function $\mathcal{T}(\mathbf{K})$ obtained from \eqref{eq:LQR1} can be expressed as:
\begin{equation}
\label{eq:LQR2}
   C(\mathbf{K}) \triangleq \|\mathcal{T}(\mathbf{K})\|^2 = \tr \left( \left( \mathbf{Q} + \mathbf{K}^{\top}\mathbf{R}\mathbf{K} \right) \bm{\Sigma}_{\mathbf{K}} \right).
\end{equation}
%\costanzo{In this formula $\|\mathcal{T}(\mathbf{K})\|$ should be squared: 
%$\|\mathcal{T}(\mathbf{K})\|^2= \tr(\cdot)$}
where $C(\mathbf{K})$ represents the cost function, and $\bm{\Sigma}_{\mathbf{K}}$ is typically referred to as the closed-loop state covariance matrix, which is the solution to the following Lyapunov equation:
\begin{equation}
\label{eq:LQR3}
   \bm{\Sigma}_{\mathbf{K}} =  \mathbf{I}_n + \left(\mathbf{A} + \mathbf{B}\mathbf{K} \right)  \bm{\Sigma}_{\mathbf{K}} \left(\mathbf{A} + \mathbf{B}\mathbf{K} \right)^{\top}.
\end{equation}
Therefore, the LQR design can be summarized as:
\begin{equation}
\label{eq:LQROptimization} 
\begin{aligned}
    \min_{\bm{\Sigma}_{\mathbf{K}} \succeq 0, \mathbf{K}} \quad & C(\mathbf{K}) = \text{Tr} \left( \left( \mathbf{Q} + \mathbf{K}^{\top}\mathbf{R}\mathbf{K} \right)\bm{\Sigma}_{\mathbf{K}} \right) \\
    \text{subject to} \quad & \bm{\Sigma}_{\mathbf{K}} =  \mathbf{I}_n + \left( \mathbf{A} + \mathbf{B}\mathbf{K} \right) \bm{\Sigma}_{\mathbf{K}} \left( \mathbf{A} + \mathbf{B}\mathbf{K} \right)^{\top}. 
\end{aligned}
\end{equation}
% \alessandro{We have not defined what $\mathbf{K}, \Sigma \succeq 0$ means. I would also use $\mathbf{X} \succeq 0$ to indicate semi-positive definite matrices.
% Mojtaba: Done!}
To directly compute the optimal feedback gain matrix $\mathbf{K}$ from~\eqref{eq:LQROptimization}, the system matrices $(\mathbf{A}, \mathbf{B})$ must be known. However, if $(\mathbf{A}, \mathbf{B})$ are unknown, it may still be possible to incorporate an identification step to estimate them. Suppose signals of length $t$ of states, inputs, noises, and successor states, which do not necessarily need to be consecutive. These signals are defined as follows:
\begin{equation}
\label{eq:data}
\begin{aligned}
    {\mathbf{X}_0} &\triangleq \begin{bmatrix}
     \mathbf{x}_{0} & \mathbf{x}_{1} & \cdots & \mathbf{x}_{t-1}
    \end{bmatrix}, \\
    {\mathbf{X}_1} &\triangleq \begin{bmatrix}
     \mathbf{x}_{1} & \mathbf{x}_{2} & \cdots & \mathbf{x}_{t}
    \end{bmatrix}, \\
    {\mathbf{U}_0} &\triangleq \begin{bmatrix}
     \mathbf{u}_{0} & \mathbf{u}_{1} & \cdots & \mathbf{u}_{t-1}
    \end{bmatrix}, \\
    {\mathbf{W}_0} &\triangleq \begin{bmatrix}
     \boldsymbol{\omega}_{0} & \boldsymbol{\omega}_{1} & \cdots & \boldsymbol{\omega}_{t-1}
    \end{bmatrix}.
\end{aligned}
\end{equation}
The input signal ${\mathbf{U}_0}$ must be \textit{sufficiently rich} to effectively represent the dynamical system described by \eqref{eq:system}. This property is commonly referred to as {\it{persistently exciting}}, and is formally defined as follows:

\begin{definition}[\cite{WILLEMS2005}]
A signal ${\mathbf{U}_0}$ is said to be persistently exciting of order $l$ when
\begin{equation}
\mathcal{U}_0 = \begin{bmatrix}
    \mathbf{u}_0 & \mathbf{u}_1 & \cdots & \mathbf{u}_{t-l} \\
    \mathbf{u}_1 & \mathbf{u}_2 & \cdots & \mathbf{u}_{t-l+1} \\
    \vdots & \vdots & \ddots & \vdots \\
    \mathbf{u}_{l-1} & \mathbf{u}_{l} & \cdots & \mathbf{u}_{t-1}
\end{bmatrix},
\end{equation}
has full rank $ml$.
\(\hfill \blacksquare\)
\label{persistently_exciting_inputs}
\end{definition}
%\alessandro{I guess that $\mathcal{U}_0$ should be ${\mathbf{U}_0}$. Also, is it a definition? If it is, isn't it an \textit{if and only if}?

%Mojtaba: we have used $\mathbf{U}_0$ in (6) for a different matrix. I guess since it is definition and not a lemma, "if" has linguistic meaning and for instance can be changed with 'when'. I changed it to avoid misunderstandings. }
The following lemma is also useful for determining the persistent excitation of a system.

\begin{lemma}[\cite{WILLEMS2005}]    \label{lemma_fundamental}
    If the system \eqref{eq:system} is controllable and ${\mathbf{U}_0}$ is persistently exciting of order $n+1$, then $\text{rank}(\mathcal{D}) = n + m$,
where
\begin{equation}
\mathcal{D} \triangleq \begin{bmatrix}
       {\mathbf{U}_0} \\
       {\mathbf{X}_0}
    \end{bmatrix}.
\end{equation}
\(\hfill \blacksquare\)
\end{lemma}

If ${\mathbf{U}_0}$ is persistently exciting, the estimates of the system matrices $(\mathbf{A}, \mathbf{B})$ can be obtained by solving the following optimization problem:
\begin{equation}
\label{eq:estimate}
    (\hat{\mathbf{A}}, \hat{\mathbf{B}}) = \min_{\mathbf{A}, \mathbf{B}} \quad  \left\| {\mathbf{X}_1} - \begin{bmatrix}
        \mathbf{B} & \mathbf{A}
    \end{bmatrix} \begin{bmatrix}
        {\mathbf{U}_0}^{\top} & {\mathbf{X}_0}^{\top}
    \end{bmatrix}^{\top} \right\|.
\end{equation}
% \alessandro{Isn't it $\begin{bmatrix}
%         {\mathbf{U}_0}^\top & {\mathbf{X}_0}^\top
%     \end{bmatrix}^{\top}$. Check carefully what is the proper transposed version.
%     Mojtaba: yes corrected.}
The LQR controller can then be designed by substituting $(\hat{\mathbf{A}}, \hat{\mathbf{B}})$, obtained from \eqref{eq:estimate}, in place of the true system matrices $(\mathbf{A}, \mathbf{B})$ in  \eqref{eq:LQROptimization}. This approach is commonly referred to as {\it{certainty-equivalence}} and is a typical strategy in {\it{indirect data-driven}} LQR design \cite{Horia2019,Sattar2022, Pilipovsky2023, Liu2024}.

Recently, several methods have been proposed in the literature to bypass the identification step in \eqref{eq:estimate} by directly leveraging the data introduced in \eqref{eq:data}. These methods are commonly referred to as \textit{direct data-driven} LQR design \cite{Tesi2020,Mohammadi2021, Tesi2021, Dorfler2022, Lopez2023, Dorfler2023, Fan2024}. From Lemma \ref{lemma_fundamental}, we know that if ${\mathbf{U}_0}$ is persistently exciting of order $n+1$, then $\operatorname{rank}(\mathcal{D}) = n + m$. Consequently, by the Rouch\'e--Capelli theorem \cite{ShafarevichRemizov2013}, there exists a matrix $\mathbf{G} \in \mathbb{R}^{t \times n}$ such that:  
\begin{equation}
    \begin{bmatrix}
        \mathbf{K} \\
        \mathbf{I}_n
    \end{bmatrix} = \mathcal{D} \mathbf{G}.
\end{equation}
Substituting into the state-space representation, we have:
\begin{equation}
\label{eq:A+BK}
    \mathbf{A}+ \mathbf{B}\mathbf{K} = \begin{bmatrix}
        \mathbf{B} & \mathbf{A}
    \end{bmatrix} \begin{bmatrix}
        \mathbf{K} \\
        \mathbf{I}_n
    \end{bmatrix} = \begin{bmatrix}
        \mathbf{B} & \mathbf{A}
    \end{bmatrix} \mathcal{D} \mathbf{G}.
\end{equation}
On the other hand, the measured data in \eqref{eq:data} must satisfy the system dynamics:
\begin{equation}
\label{eq:datadynamic}
    {\mathbf{X}_1} = \mathbf{A} {\mathbf{X}_0} + \mathbf{B} {\mathbf{U}_0} + {\mathbf{W}_0}.
\end{equation}
By substituting the definition of $\mathcal{D}$ from Lemma \ref{lemma_fundamental} into \eqref{eq:A+BK} and incorporating \eqref{eq:datadynamic}, we obtain:
\begin{equation}
\label{eq:A+BK_2}
    \mathbf{A}+ \mathbf{B}\mathbf{K} = ({\mathbf{X}_1} - {\mathbf{W}_0}) \mathbf{G}.
\end{equation}
Since ${\mathbf{W}_0}$ is unknown and cannot be directly accounted for, we approximate $\mathbf{A}+ \mathbf{B}\mathbf{K}$ with ${\mathbf{X}_1} \mathbf{G}$ and $\mathbf{K}$ with ${\mathbf{U}_0} \mathbf{G}$ in the LQR optimization problem. This leads to the following direct data-driven LQR optimization formulation:
\begin{equation}
\label{eq:LQRDDOptimization}
\begin{aligned}
    \min_{ \bm{\Sigma}_{\mathbf{K}} \succeq 0, \mathbf{G}} \quad & C(\mathbf{G})=\tr \left( \left( \mathbf{Q} + \mathbf{G}^{\top} {\mathbf{U}_0}^{\top} \mathbf{R}{\mathbf{U}_0}\mathbf{G} \right) \bm{\Sigma}_{\mathbf{K}} \right) \\
    \text{subject to} \quad & \bm{\Sigma}_{\mathbf{K}} =  \mathbf{I}_n + {\mathbf{X}_1}\mathbf{G} \bm{\Sigma}_{\mathbf{K}} \mathbf{G}^{\top} {\mathbf{X}_1}^{\top},\\
    & {\mathbf{X}_0}\mathbf{G} = \mathbf{I}_n.
\end{aligned}
\end{equation}
The optimal feedback gain is then given by $\mathbf{K} = {\mathbf{U}_0}\mathbf{G}$.

In \cite{zhao2024data}, the authors introduce an alternative policy parametrization based on the sample covariance of the data, defined as:
\begin{equation}
\label{eq:covariance_parametrization}
\Phi \triangleq \frac{1}{t} \mathcal{D} \mathcal{D}^{\top} = \begin{bmatrix}
    {\mathbf{U}_0} \mathcal{D}^{\top} / t \\
    {\mathbf{X}_0} \mathcal{D}^{\top} / t 
\end{bmatrix} = \begin{bmatrix}
    \overline{{\mathbf{U}}}_0 \\
    \overline{{\mathbf{X}}}_0
\end{bmatrix}.
\end{equation}
% \alessandro{I would re-write $\overline{{\mathbf{U}}}_0$ and $\overline{{\mathbf{X}}}_0$ as $\overline{{\mathbf{U}}}_0$ and $\overline{{\mathbf{X}}}_0$.}
Defining $\mathbf{V} \in \mathbb{R}^{(n+m) \times n}$ as the solution to:
\begin{equation}
\label{eq:phivi}
\begin{bmatrix}
        \mathbf{K} \\
        \mathbf{I}_n
    \end{bmatrix} = \Phi \mathbf{V},
\end{equation}
and following steps similar to \eqref{eq:A+BK}--\eqref{eq:A+BK_2}, the data-driven LQR optimization problem can be reformulated as:
\begin{equation}
\label{eq:LQRDDCPOptimization} 
\begin{aligned}
    \min_{\bm{\Sigma}_{\mathbf{V}} \succeq 0, \mathbf{V}} \quad & C(\mathbf{V}) = \tr \left( \left( \mathbf{Q} + \mathbf{V}^{\top} \overline{{\mathbf{U}}}_0^{\top} \mathbf{R}\overline{{\mathbf{U}}}_0\mathbf{V} \right) \bm{\Sigma}_{\mathbf{V}} \right) \\
    \text{subject to} \quad & \bm{\Sigma}_{\mathbf{V}} =  \mathbf{I}_n + \overline{{\mathbf{X}}}_1\mathbf{V} \bm{\Sigma}_{\mathbf{V}} \mathbf{V}^{\top} \overline{{\mathbf{X}}}_1^{\top},\\
    & \overline{{\mathbf{X}}}_0\mathbf{V} = \mathbf{I}_n,
\end{aligned}
\end{equation}
where $\overline{{\mathbf{X}}}_1 = {\mathbf{X}_1} \mathcal{D}^{\top} / t $.  Since the dimension of $\mathbf{V}$ is independent of the number of samples, $t$, this formulation is particularly advantageous in adaptive design strategies where the sample size grows linearly.

In the DeePO algorithm, starting from an initial feasible solution $\mathbf{K}_t$, the feedback gain evolves iteratively via a gradient descent approach to reach the optimal solution $\mathbf{K}^{*}$.
% \begin{equation}
%     \mathbf{K}_{i+1} = \mathbf{K}_{i} - \eta \widehat{\nabla C},
% \end{equation}
% where $\eta$ is the learning rate and $\widehat{\nabla C}$ is a gradient estimate of the objective function obtained through zeroth-order information. 
The following lemma provides a methodology for computing $\widehat{\nabla_{\mathbf{V}} C}$ solely based on data.
% \costanzo{The function
% $C(\cdot)$  has been defined in \eqref{eq:LQR2} and in \eqref{eq:LQROptimization} as a function of $\mathbf{K}$, 
% in \eqref{eq:LQRDDCPOptimization} as a function of $\mathbf{}{G}$ 
% and in \eqref{eq:LQRDDOptimization} as a function of $\mathbf{V}$.
% As a consequence it is not obvious what is the estimated gradient of $C$ used in this equation (although $C(\mathbf{K})$ seems the most likely one at this stage).
% However, looking at the gradient in the next Lemma 2, one may think that it is the gradient used of $C(\mathbf{V})$ defined in \eqref{eq:LQRDDOptimization}.
% Maybe a subscript can help  
% (e.g., $\widehat{\nabla_{\mathbf{K}} C}$ in the previous equation and $\widehat{\nabla_{\mathbf{V}} C}$ in the following Lemma). }

\begin{lemma}[\cite{zhao2024data}]
    Let $\mathbf{P}_{\mathbf{V}}$ be the unique solution for the Lyapunov equation
    $$\mathbf{P}_{\mathbf{V}} = {\mathbf{Q}} + \mathbf{V}^{\top} \overline{{\mathbf{U}}}_0^{\top} \mathbf{R}\overline{{\mathbf{U}}}_0\mathbf{V} + \mathbf{V}^{\top} \overline{{\mathbf{X}}}_1^{\top} \mathbf{P}_{\mathbf{V}}\overline{{\mathbf{X}}}_1\mathbf{V},$$
    \label{lemma_gradient}
    then
    $$\widehat{\nabla_{\mathbf{V}} C} = 2 \left(\overline{{\mathbf{U}}}_0^{\top} \mathbf{R}\overline{{\mathbf{U}}}_0 + \overline{{\mathbf{X}}}_1^{\top} \mathbf{P}_{\mathbf{V}}\overline{{\mathbf{X}}}_1 \right)\mathbf{V}\Sigma_{\mathbf{V}}.$$
    \(\hfill \blacksquare\)
\end{lemma}

Algorithm \ref{Algo:DeePO} outlines the steps required to execute the DeePO algorithm.

\begin{algorithm}[ht] 
\small
\caption{{Data-Enabled Policy Optimization (DeePO) \cite{zhao2024data}} \label{Algo:DeePO}}
\def\negsp{\vspace{-5pt}}
\def\negup{\vspace{-7pt}}
\def\negin{\vspace{-3pt}}
\def\posin{0.25cm}
\setstretch{1.2}
 \begin{algorithmic}[section]
 \Require $\mathbf{U}_{0}$, ${\mathbf{X}_0}$, ${\mathbf{X}_1}$, $\mathbf{K}_t$, a probing noise signal $\{\mathbf{e}\}$, and a learning rate $\eta$.
 \Statex \hspace{-0.5cm} {\bf{Start}}
 \Statex \hspace{-0.5cm} {$i=t$.}
 \Statex \hspace{-0.3cm}{\bf{while the stop criterion is not satisfied, do:}}
 \Statex \hspace{+0.0cm} Apply ${{\mathbf{u}}}_i = \mathbf{K}_i\mathbf{x}_i + \mathbf{e}_i$ and observe ${{\mathbf{x}}}_{i+1}$.
 \Statex \hspace{+0.0cm} Update ${\mathbf{X}_0}$ by ${\mathbf{X}_0} = [{\mathbf{X}_0}, \mathbf{x}_i]$.
 \Statex \hspace{+0.0cm} Update ${\mathbf{X}_1}$ by ${\mathbf{X}_1} = [{\mathbf{X}_1}, \mathbf{x}_{i+1}]$.
 \Statex \hspace{+0.0cm} Update ${\mathbf{U}_0}$ by ${\mathbf{U}_0} = [{\mathbf{U}_0}, \mathbf{u}_i]$.
 \Statex \hspace{+0.0cm} ${{\mathbf{V}}}_{i+1} = \Phi_{i+1}^{-1} \begin{bmatrix}
     \mathbf{K}_i \\ \mathbf{I}_n
 \end{bmatrix}$.
 \Statex \hspace{+0.0cm} ${{\mathbf{V}}}^{\prime}_{i+1} = {{\mathbf{V}}}_{i+1} - \eta \Pi_{\overline{{\mathbf{X}}}_0} \widehat{\nabla_{\mathbf{V}} C}$.
 \Statex \hspace{+0.0cm} Update the control gain by $\mathbf{K}_{i+1} = {\overline{{\mathbf{U}}}_0}{{\mathbf{V}}}^{\prime}_{i+1} $.
 \Statex \hspace{+0.0cm} {$i=i+1$.}
 \Statex \hspace{-0.3cm} {\bf{End while}}
 \Statex \hspace{-0.5cm} {\bf{End}}
 \end{algorithmic}
\end{algorithm}
% \costanzo{Here the projection operator $\Pi_{\overline{{\mathbf{X}}}_0}$ has not been defined nor justified.}
In summary, to ensure the convergence of Algorithm~\ref{Algo:DeePO}, the optimization problem~\eqref{eq:LQRDDCPOptimization} must be feasible, ${\mathbf{U}_0}$ must be persistently exciting, and both $\boldsymbol{\omega}_k$ and $\mathbf{x}_k$ must remain bounded to maintain stability.
% \costanzo{The reason for boundedness of $\boldsymbol{\omega}_k$ and $\mathbf{x}_k$ has not been explained yet.
% The reader will discover that the boundedness of $\mathbf{x}_k$ is needed in the next Lemma 3. 
% Is there another motivation? What about $\boldsymbol{\omega}_k$?}

\subsection{Why is probing noise added to $\mathbf{u}_i$ in Algorithm~\ref{Algo:DeePO}?}

To begin, we present a lemma demonstrating that state feedback control of the form $\mathbf{u}_k = \mathbf{K}\mathbf{x}_k$ is incapable of generating a persistently exciting sequence ${\mathbf{U}_0}$.

\begin{lemma}
State feedback control of the form $\mathbf{u}_k = \mathbf{K}\mathbf{x}_k$, where $\mathbf{K}$ is a gain matrix and $\mathbf{x}_k$ is the state vector at time step $k$, cannot generate a persistently exciting sequence ${\mathbf{U}_0}$.
\label{lemma:fixedfeedback}
\end{lemma}

\begin{proof}
From Lemma \ref{lemma_fundamental}, we construct the matrix $\mathcal{D}$ as follows:
\begin{equation}
    \mathcal{D} = 
    \begin{bmatrix}
        \sum_{j=1}^{n} \mathbf{k}^{1,j}\mathbf{x}_0^j & \sum_{j=1}^{n} \mathbf{k}^{1,j}\mathbf{x}_1^j & \cdots & \sum_{j=1}^{n} \mathbf{k}^{1,j}\mathbf{x}_{t-1}^j \\
        \vdots & \vdots & \ddots & \vdots \\
        \sum_{j=1}^{n} \mathbf{k}^{m,j}\mathbf{x}_0^j & \sum_{j=1}^{n} \mathbf{k}^{m,j}\mathbf{x}_1^j & \cdots & \sum_{j=1}^{n} \mathbf{k}^{m,j}\mathbf{x}_{t-1}^j \\
        \mathbf{x}_0^1 & \mathbf{x}_1^1 & \cdots & \mathbf{x}_{t-1}^1 \\
        \vdots & \vdots & \ddots & \vdots \\
        \mathbf{x}_0^n & \mathbf{x}_1^n & \cdots & \mathbf{x}_{t-1}^n
    \end{bmatrix},
\end{equation}
where first $m$ rows of $\mathcal{D}$ can be expressed as linear combinations of the last $n$ rows. Consequently, $\text{rank}(\mathcal{D}) = n$. From Lemma \ref{lemma_fundamental}, we observe that the condition $\text{rank}(\mathcal{D}) \neq n + m$ implies that the sequence ${\mathbf{U}_0}$ is not persistently exciting. This completes the proof.
\end{proof}

In addition, the following lemma provides a more transparent perspective on the subject.

\begin{lemma}
\label{lemma:rankphi}
Let $\mathcal{D}$ be defined as the horizontal concatenation of two matrices,
$\mathcal{D} = [\mathcal{D}_{1}, \mathcal{D}_{2}]$,

where:
\begin{itemize}
    \item $\mathcal{D}_{1} \in \mathbb{R}^{(n+m) \times t_1}$ with $\text{rank}(\mathcal{D}_{1}) = n + m$, and
    \item $\mathcal{D}_{2} \in \mathbb{R}^{(n+m) \times t_2}$ with $\text{rank}(\mathcal{D}_{2}) < n + m$.
\end{itemize}
If $t_2 \to \infty$, then
$\lim_{t_2 \to \infty} \underline{\sigma}(\Phi) \to 0$.

%where $\underline{\sigma}(\Phi)$ denotes the smallest singular value of $\Phi$.
\end{lemma}

\begin{proof}
First, notice that $\Phi$ defined in \eqref{eq:covariance_parametrization} is only  a function of $t_2$, since $t_1$ is fixed:
\begin{equation}
\Phi(t_2) = \frac{1}{t_1 + t_2} 
(\mathcal{D}_{1}\mathcal{D}_{1}^{\top} + \mathcal{D}_{2}\mathcal{D}_{2}^{\top}).
\end{equation}
Since $\mathcal{D}_{2}$ is not full row rank for all $t_2$, there exists a non-zero vector $\mathbf{a}\in \mathbb{R}^{n+m}$ such that $\mathbf{a}^\top \mathcal{D}_{2}=0$.
Thus
\begin{equation}
\begin{aligned}    
    \mathbf{a}^\top \Phi(t_2)\mathbf{a} & = 
    \frac{1}{t_1 + t_2} (\mathbf{a}^\top \mathcal{D}_{1}\mathcal{D}_{1}^{\top} \mathbf{a}
    +\mathbf{a}^\top \mathcal{D}_{2}\mathcal{D}_{2}^{\top} \mathbf{a})      \\
    & =   \frac{1}{t_1 + t_2} \mathbf{a}^\top \mathcal{D}_{1}\mathcal{D}_{1}^{\top} \mathbf{a}.
\end{aligned}
\end{equation}
From this, it is clear that 
\begin{equation} \label{eq:limitaPhia}
    \lim_{t_2\to\infty }\mathbf{a}^\top \Phi(t_2)\mathbf{a} = 0.
\end{equation}
Since $\underline{\sigma}(\Phi(t_2))$ is such that
\begin{equation}
\underline{\sigma}(\Phi(t_2)) \Vert \mathbf{a} \Vert^2 \le \mathbf{a}^\top \Phi(t_2)\mathbf{a},\quad \forall t_2\in\mathbb{N},
\end{equation}
then the limit \eqref{eq:limitaPhia} implies
\[
\lim_{t_2 \to \infty} \underline{\sigma}\big(\Phi(t_2)\big) \to 0.
\]
This concludes the proof.
\end{proof}

\noindent
The results from Lemmas \ref{lemma:fixedfeedback} and \ref{lemma:rankphi} reveal critical limitations for applying conventional control laws, such as $\mathbf{u}_i = \mathbf{K}_i \mathbf{x}_i$, to the system:
\begin{itemize}
    \item As $\mathbf{x}_i \to 0$, the columns of $\mathcal{D}$ also approach zero,
    and $\mathcal{D}$ can be partitioned such that the second matrix is zero. By Lemma \ref{lemma:rankphi}, $\lim_{t_2 \to \infty} \underline{\sigma}(\Phi) \to 0$, jeopardizing the full rank of $\Phi$ needed to ensure \eqref{eq:phivi}.
    
    \item As $\mathbf{K}_i \to \mathbf{K}$, Lemma \ref{lemma:fixedfeedback} indicates that adding columns to $\mathcal{D}$ is equivalent to appending a singular matrix to $\mathcal{D}_{1}$, pushing $\Phi$ toward singularity.
\end{itemize}

To address this undesired outcome, which compromises DeePO's performance, the authors in~\cite{zhao2024data} propose adding a probing noise to the input. This noise ensures that the input remains sufficiently rich, thereby preserving the full rank of $\Phi$ and maintaining persistent excitation.

\subsection{Undesired State Perturbations by Adding Probing Noise to $\mathbf{u}_i$}

In asymptotically stable LTI systems, the system naturally drives the state $\mathbf{x}_i$ towards equilibrium as time progresses. 
However, when probing noise is added to the input signal to maintain persistent excitation, the noise introduces high-frequency components into the control input. These high-frequency components can interact with the feedback dynamics, causing rapid oscillations or fluctuations in the control signal and, consequently, the system state. States perturbations is particularly problematic in practical implementations, as it can lead to actuator wear, increased energy consumption, and degraded overall system performance.

\section{Perturbations-free DeePO (PFDeePO)}
\label{sec:Algos}

As discussed in Section~\ref{sec:DeePO}, the convergence of $\mathbf{x}_i \to 0$ and $\mathbf{K}_i \to \mathbf{K}$ can compromise the full rank of $\Phi$. In Algorithm~\ref{algo:state perturbationsfreeDeePO}, we present our proposed method, PFDeePO, which serves as an alternative to DeePO, ensuring the full rank of $\Phi$ is preserved while avoiding performance degradation caused by state perturbations.
% \costanzo{It seems to me that the use of random multiplicative coefficients $\nu_i$ can avoid that first $m$ rows of $\mathcal{D}$ asymptotically tend to be linear combinations of the last $n$ rows. 
% However, this solution does not avoid that the state exponentially goes to zero, and therefore the asymptotic rank-loss of $\Phi$ due to this reason appears to be unavoidable. Am I missing something in your reasoning?}

\begin{algorithm}[ht] 
\small
\caption{ Perturbations-free DeePO (PFDeePO) }
\label{algo:state perturbationsfreeDeePO}
\def\negsp{\vspace{-5pt}}
\def\negup{\vspace{-7pt}}
\def\negin{\vspace{-3pt}}
\def\posin{0.25cm}
\setstretch{1.2}
 \begin{algorithmic}[section]
 \Require $\mathbf{U}_{0}$, ${\mathbf{X}_0}$, ${\mathbf{X}_1}$, $\mathbf{K}_t$, $\gamma > 0$, $\delta >0$, and $\eta>0$.
 \Statex \hspace{-0.5cm} {\bf{Start}}
 \Statex \hspace{-0.5cm} {$i=t$.}
 \Statex \hspace{-0.5cm} {$\Delta \mathbf{K}= (\delta+1) \cdot \mathbf{1}_{m \times n}$.}
 \Statex \hspace{-0.3cm}{\bf{while the stop criterion is not satisfied, do:}}
 \Statex \hspace{+0.0cm}{\bf{if $\|\Delta \mathbf{K}\| > \delta$ or $\|\mathbf{x}_i \|\le \gamma$}}
 \Statex \hspace{+0.3cm} Apply ${{\mathbf{u}}}_i = \mathbf{K}_i\mathbf{x}_i$ and observe ${{\mathbf{x}}}_{i+1}$.
 \Statex \hspace{+0.0cm}{\bf{else}}
 \Statex \hspace{+0.3cm} Find $\overline{v}$ and $\underline{v}$ as in Theorem \ref{Thm:mainTheo} for $\mathbf{K}_i$.
 \Statex \hspace{+0.3cm} Randomly select  $\underline{v} \le v_i \le \overline{v} $.
 \Statex \hspace{+0.3cm} Apply ${{\mathbf{u}}}_i = v_i\mathbf{K}_i\mathbf{x}_i$ and observe ${{\mathbf{x}}}_{i+1}$.
 \Statex \hspace{+0.0cm}{\bf{End if}}
 \Statex \hspace{+0.0cm}{\bf{if $\|\mathbf{x}_i \|> \gamma$}}
 \Statex \hspace{+0.3cm} Update ${\mathbf{X}_0}$ by ${\mathbf{X}_0} = [{\mathbf{X}_0}, \mathbf{x}_i]$.
 \Statex \hspace{+0.3cm} Update ${\mathbf{X}_1}$ by ${\mathbf{X}_1} = [{\mathbf{X}_1}, \mathbf{x}_{i+1}]$.
 \Statex \hspace{+0.3cm} Update ${\mathbf{U}_0}$ by ${\mathbf{U}_0} = [{\mathbf{U}_0}, \mathbf{u}_i]$.
 \Statex \hspace{+0.3cm} ${{\mathbf{V}}}_{i+1} = \Phi_{i+1}^{-1} \begin{bmatrix}
     \mathbf{K}_i \\ \mathbf{I}_n
 \end{bmatrix}$.
 \Statex \hspace{+0.3cm} ${{\mathbf{V}}}^{\prime}_{i+1} = {{\mathbf{V}}}_{i+1} - \eta \Pi_{\overline{{\mathbf{X}}}_0} \widehat{\nabla_{\mathbf{V}} C}$.
 \Statex \hspace{+0.3cm} Update the control gain by $\mathbf{K}_{i+1} = {\overline{{\mathbf{U}}}_0}{{\mathbf{V}}}^{\prime}_{i+1} $.
 \Statex \hspace{+0.0cm}{\bf{else}}
 \Statex \hspace{+0.3cm} Update the control gain by $\mathbf{K}_{i+1} = \mathbf{K}_{i} $.
 \Statex \hspace{+0.0cm}{\bf{End if}}
 \Statex \hspace{+0.0cm} {$\Delta \mathbf{K}= \mathbf{K}_{i+1} - \mathbf{K}_{i}$}.
 \Statex \hspace{+0.0cm} {$i=i+1$.}
 \Statex \hspace{-0.3cm}{\bf{End while}}
 \Statex \hspace{-0.5cm}{\bf{End}}
 \end{algorithmic}
\end{algorithm}

The main idea behind PFDeePO is to prevent conditions that may compromise the full rank of $\Phi$. To achieve this, we introduce two additional positive scalars, $\gamma > 0$ and $\delta > 0$. 

At each iteration, the first ``if-else'' block ensures that arbitrarily scaling the control signal by a random gain, $v_i$, where $\underline{v} \leq v_i \leq \overline{v}$, occurs only when necessary. Specifically, this scaling is applied only if the control gain, $\mathbf{K}_i$, has already converged to its optimal value, yet the system states remain relatively far from the equilibrium point.

Next, in the second ``if-else" block, we evaluate the convergence of the states. The condition $\|\mathbf{x}_i\| < \gamma$ represents a scenario where the states are close to the equilibrium point, causing the control signal to approach zero. In this situation, the data gathered is not sufficiently informative to update the controller. Conversely, if the states have already reached near-equilibrium, further efforts to improve the control law are unnecessary, as the states have effectively converged, and any additional effort is unlikely to yield meaningful improvements in classical control performance metrics.

To formalize PFDeePO, we first need to ensure that the matrix \(\Phi_i\) retains full rank when Algorithm~\ref{algo:state perturbationsfreeDeePO} is applied.
% \costanzo{I see that \(\Phi_i\), for any finite $i$, remains full-rank, but this does not forbid the fact that asymptotically the smallest singular value goes to zero, due to the fact that that both the state $\mathbf{x_i}$ and input $\mathbf{u_i}$ go to zero.
% Conversely, I understand that if $\mathbf{x_i}$ is practically zero (below the threshold $\gamma$), there is no need to update the gain $\mathbf{K}_i$, and therefore the full rank property for \(\Phi_i\) in this situation is not so important.
% I am not sure this is the spirit of this proposal.}

\begin{theorem}
Let \(\Phi_i\) be the matrix constructed at time step \(i\) during the execution of PFDeePO. By implementing Algorithm~\ref{algo:state perturbationsfreeDeePO}, the matrix \(\Phi_i\) attains full rank, i.e., \(\text{rank}(\Phi_i) = n + m\), and as a result, $\underline{\sigma}(\Phi_i) > 0$.
\end{theorem}
\begin{proof}
From \eqref{eq:covariance_parametrization} we see that \(\text{rank}(\Phi_i) = \text{rank}(\mathcal{D})\). 
We proceed by contradiction. Assume that \(\text{rank}(\mathcal{D}) < n + m\). This implies the existence of scalars \(\lambda_i, \mu_j \in \mathbb{R}\) such that:
\begin{equation}
    \sum_{i=1}^{n} \lambda_i {\mathbf{X}_0}^i = \sum_{j=1}^{m} \mu_j {\mathbf{U}_0}^j,
    \label{eq:theorem1_1}
\end{equation}
where \({\mathbf{X}_0}^i\) denotes the \(i\)-th row of \({\mathbf{X}_0}\), and \({\mathbf{U}_0}^j\) denotes the \(j\)-th row of \({\mathbf{U}_0}\). 
Note that the second if-else block, avoids presence of all zeros columns in $\mathbf{X}_0$ and $\mathbf{U}_0$ and 
recall that at each time step $k$, \({\mathbf{U}_0}^{j,k} = \sum_{i=1}^{n} \mathbf{k}_k^{j,i} \mathbf{x}_k^i\), which holds for all \(k = 0, \dots, t-1\), we substitute into \eqref{eq:theorem1_1} to obtain:
\begin{equation}
    \sum_{i=1}^{n} \lambda_i \mathbf{x}_k^i = \sum_{j=1}^{m} \sum_{i=1}^{n} \mu_j \mathbf{k}_k^{j,i} \mathbf{x}_k^i.
    \label{eq:theorem1_2}
\end{equation}
For \eqref{eq:theorem1_2} to hold for all \(k = 0, \dots, t-1\) and \(i = 1, \dots, n\), we must have:
\begin{equation}
    \lambda_i = \sum_{j=1}^{m} \mu_j \mathbf{k}_k^{j,i}.
    \label{eq:theorem1_3}
\end{equation}
However, this condition \eqref{eq:theorem1_3} is violated under both cases defined by Algorithm~\ref{algo:state perturbationsfreeDeePO}:
\begin{itemize}
    \item When \(\mathbf{K}_k \neq \mathbf{K}_{k-1}\), the matrices \(\mathbf{K}_k\) are distinct, and it is impossible for \(\lambda_i\) to be expressed as a consistent linear combination of \(\mu_j \mathbf{k}_k^{j,i}\). In other words, since the \(i\)-th column of the controller changes at each time step, the value of \(\lambda_i\) obtained at time step \(k\) as a linear combination of \(\mu_j \mathbf{k}_k^{j,i}\) will differ from the values obtained at other time steps and it is not feasible to find a consistent \(\lambda_i\) for all \(k = 0, \dots, t-1\).

    \item When \(\mathbf{K}_k = v_k \mathbf{K}\), where \(v_i\) is randomly sampled from an arbitrary distribution, the randomness in \(v_k\) ensures that $\lambda_i$ will be different value at each time step in $\lambda_i = \sum_{j=1}^{m} v_k\mu_j \mathbf{k}^{j,i}$.
\end{itemize}

Thus, the assumption that \(\text{rank}(\mathcal{D}) < n + m\) leads to a contradiction. Therefore, we conclude that:
\[
\text{rank}(\mathcal{D}) = n + m.
\]
\end{proof}

% \costanzo{I think that the notations should be improved to better convey the ideas. 
% First, I note that the matrix $\Phi$ with subscript $i$ has not been formally defined. 
% Even in the Algorithms 1 and 2, there is not a formula for its computation.
% The matrix $\Phi$ has been defined in \eqref{eq:covariance_parametrization} without any subscript.
% Maybe it should be defined as $\Phi_t$, for future purposes? 
% Consequently, the matrix $\mathcal{D}$ should have the same subscript $t$.
% It would make sense to have $\Phi_i = \frac{1}{i}\mathcal{D}_i\mathcal{D}_i^\top$.
% Note also, that the initial definitions of ${\mathbf{X}_0}$, ${\mathbf{X}_1}$, ${\mathbf{U}_0}$, where the final time $t$ does not appear in the notation, is good for a one-shot computation of the LQR gain, but is not very suitable for the on-line adaptation of the gain.
% Indeed, both $\Phi$ and $\mathcal{D}$ are defined as functions of ${\mathbf{X}_0}$ and ${\mathbf{U}_0}$, for given $t$, that does not appear in the notation.}
% \vitt{I agree that the notation could be improved due to the 'real-time' scenario. A sort of De Persis/Tesi (taken from standard notation in subspace identification) would be better: $X_{0,t}, X_{1,t}$ for (6) and maybe $\mathcal{U}_{0,l,t-l+1}$ in (7). Similarly, one could define $\mathcal{D}_{0,t}$. In subspace notation first index is starting time, and other refer to length, not to final/current time (but they can be adapted if indexing on the final/current time is better). Problematic change of notation with close deadline, though.}

Another critical aspect to address is ensuring that randomly scaling the control signal by \(v_i\) does not compromise the closed-loop stability. First recall that from Lemma 1 in \cite{zhao2024data}, the converged control gain $\mathbf{K}$ obtained by DeePO is equivalent to indirect LQR solution. Therefore, 
\begin{equation}
    \mathbf{K} = \left(\hat{\mathbf{B}}^{\top}\mathbf{H}\hat{\mathbf{B}} + \mathbf{R} \right)^{-1}\hat{\mathbf{B}}^{\top}\mathbf{H}\hat{\mathbf{A}},
    \label{eq:K}
\end{equation}
where $\mathbf{H}$ is the solution of the following Discrete Algebraic Riccati Equation (DARE)
\begin{equation}
    \mathbf{H} = \hat{\mathbf{A}}^{\top}\mathbf{H}\hat{\mathbf{A}} + \mathbf{Q} - \hat{\mathbf{A}}^{\top}\mathbf{H}\hat{\mathbf{B}} \left(\hat{\mathbf{B}}^{\top}\mathbf{H}\hat{\mathbf{B}} + \mathbf{R} \right)^{-1}\hat{\mathbf{B}}^{\top}\mathbf{H}\hat{\mathbf{A}}.
    \label{eq:Riccati}
\end{equation}
and $\mathbf{Q} \in \mathbb{R}^{n\times n}$ and  $\mathbf{R}\in\mathbb{R}^{m\times m}$ are given symmetric positive definite matrices.
In the following, we establish the conditions that \(\overline{v}\) and \(\underline{v}\) must satisfy to guarantee stability in Algorithm \ref{algo:state perturbationsfreeDeePO}. We begin by the following lemma.
 
\begin{lemma} \label{Lem:QKGKpos}
Consider matrices $\hat{\mathbf{B}}\in\mathbb{R}^{n\times m}$ and $\mathbf{K}\in\mathbb{R}^{m\times n}$,
and symmetric positive definite matrices $\mathbf{Q}\in\mathbb{R}^{n\times n}$ and $\mathbf{R}\in\mathbb{R}^{m\times m}$.
Then, there exists an interval $\mathcal{V} =[\underline{v}, \overline{v}]$, where $0<\underline{v} < 1 < \overline{v}$, such that
\begin{equation} \label{eq:ineqQKK} 
    \mathbf{Q}-\mathbf{K}^{\top}\big( (v-1)^2 \hat{\mathbf{B}}^{\top} \mathbf{H}\hat{\mathbf{B}} + (1-2v)\mathbf{R}\big)\mathbf{K} \succeq 0, \quad \forall v \in [\underline{v}, \overline{v}].
\end{equation}

 \end{lemma}
\begin{proof}
Note first that for $v=1$, we have
\begin{equation} \label{eq:ineqQKKB} 
\begin{aligned}
    \Big( \mathbf{Q} - \mathbf{K}^{\top}\big( (v-1)^2 \hat{\mathbf{B}}^{\top} \mathbf{H}\hat{\mathbf{B}} + (1-2v)\mathbf{R}\big)\mathbf{K}\Big)_{v=1}\\
    =\mathbf{Q}+\mathbf{K}^{\top}\mathbf{R}\mathbf{K} \succ 0.
    \end{aligned}
\end{equation}

Since $\mathbf{Q} \succ 0$ by assumption and $\mathbf{K}^{\top}\mathbf{R}\mathbf{K} \succeq 0$, the matrix in the left-hand side of \eqref{eq:ineqQKK} is positive definite at  $v=1$. Recall that the eigenvalues of a square matrix, whose components depend continuously on a parameter, are also continuous functions of that parameter. As a result, there exists an open neighborhood $(\underline{v},\overline{v})$ around $v = 1$ where all eigenvalues of the matrix remain strictly positive, ensuring that the matrix is positive definite. 

At the endpoints $\underline{v}$ and $\overline{v}$, the matrix becomes positive semidefinite, meaning some of its eigenvalues reach zero. Consequently, the inequality \eqref{eq:ineqQKK} holds for all $v \in [\underline{v},\overline{v}]$, completing the proof.
\end{proof}

\begin{remark}
In Lemma~\ref{Lem:QKGKpos}, we show that there exists an interval 
$\mathcal{V} = [\underline{v}, \overline{v}]$ such that \eqref{eq:ineqQKK} 
holds for all $v \in \mathcal{V}$. A practical approach to determine 
$\mathcal{V}$ is to start with a small interval around $1$ and iteratively 
narrow it, checking Lemma~\ref{Lem:QKGKpos} at each step, until the inequality 
is satisfied. \hfill$\square$
\end{remark}

% \begin{corollary}
% \label{col:lem1}
%     In a particular case of Lemma \ref{Lem:QKGKpos} , in which $m=1$, so that both $\mathbf{B}$ and $\mathbf{K}^{\top}$ are column vectors and $\mathbf{R}$ and $\mathbf{B}^{\top} \mathbf{H}\mathbf{B}$ are scalar, the following bounds for $\underline{v}$ and $\overline{v}$ hold true
% \begin{equation} \label{eq:vminvmax}
%  \begin{aligned}
%      \underline{v} & \le 1+\frac{\mathbf{R}}{\mathbf{B}^{\top} \mathbf{H} \mathbf{B}}-\Delta,\\
%      \overline{v} & \succeq 1+\frac{\mathbf{R}}{\mathbf{B}^{\top} \mathbf{H} \mathbf{B}}+\Delta,\\
%      \text{where} &\\
%      \Delta =&  \frac{\mathbf{R}}{\mathbf{B}^{\top} \mathbf{H} \mathbf{B}}\sqrt{1+ \frac{\mathbf{B}^{\top} \mathbf{H} \mathbf{B}}{\mathbf{R}}\Big( 1+ \frac{q_{\min}}{\mathbf{R}\Vert\mathbf{K}\Vert^2} \Big)}
%  \end{aligned}
%  \end{equation}
%  where $q_{\min} = \lambda_{\min}(\mathbf{Q})$ .
% \end{corollary}
% \begin{proof}
%     See Appendix.
% \end{proof}
Now, we are ready to present our main theorem.

\begin{theorem} \label{Thm:mainTheo}
Consider a system controlled by Algorithm \ref{algo:state perturbationsfreeDeePO}.  
Let $\mathbf{Q} \in \mathbb{R}^{n\times n}$ and $\mathbf{R} \in \mathbb{R}^{m\times m}$ be given symmetric and positive definite matrices, and let $\beta \in (0,1)$.  

Let's represent the converged control gain by the matrix $\mathbf{K} \in \mathbb{R}^{m\times n}$, since $\mathbf{K}$ is equal to a certainly equivalent LQR gain, we have
\begin{equation} \label{eq:defK}
    \mathbf{K} = (\mathbf{R} + \hat{\mathbf{B}}^{\top} \mathbf{H} \hat{\mathbf{B}})^{-1} \hat{\mathbf{B}}^{\top} \mathbf{H} \hat{\mathbf{A}},
\end{equation}
where $\mathbf{H} \in \mathbb{R}^{n\times n}$ is the unique solution of the modified DARE:
\begin{equation} \label{eq:DAREbeta}
    \hat{\mathbf{A}}^{\top} \mathbf{H} \hat{\mathbf{A}} - \beta^2 \mathbf{H} 
    - \hat{\mathbf{A}}^{\top} \mathbf{H} \hat{\mathbf{B}} 
    (\mathbf{R} + \hat{\mathbf{B}}^{\top} \mathbf{H} \hat{\mathbf{B}})^{-1} 
    \hat{\mathbf{B}}^{\top} \mathbf{H} \hat{\mathbf{A}} 
    + \mathbf{Q} = \mathbf{0}_{n \times n}.
\end{equation}

Consider the interval $[\underline{v},\overline{v}]$, as defined in Lemma \ref{Lem:QKGKpos}, such that the inequality \eqref{eq:ineqQKK} holds.  
Then, the time-varying state feedback control law,
$\mathbf{u}_k = -v_k \mathbf{K} \mathbf{x}_k$,
where $v_k$ is any sequence taking values in the interval $[\underline{v},\overline{v}]$, ensures that the origin of the closed-loop system  
\begin{equation} \label{eq:DLTSysClosed}
    \mathbf{x}_{k+1} = (\hat{\mathbf{A}} - v_k \hat{\mathbf{B}} \mathbf{K}) \mathbf{x}_k, \quad k = 0,1,\dots
\end{equation}
is exponentially stable. Specifically, the state satisfies the bound  
\begin{equation} \label{eq:expbound}
    \Vert \mathbf{x}_k\Vert \leq \beta^{k} \sqrt{\frac{h_{\max}}{h_{\min}}} \Vert \mathbf{x}_0\Vert, \quad k = 0,1,\dots,
\end{equation}
where  $h_{\max} = \lambda_{\max}(\mathbf{H})$, and $h_{\min} = \lambda_{\min}(\mathbf{H})$.

\end{theorem}

\begin{proof}
For a compact notation, the gain $\mathbf{K}$ defined in \eqref{eq:defK} will be written as
\begin{equation} \label{eq:defKS}
    \mathbf{K}=\mathbf{S}^{-1}\hat{\mathbf{B}}^{\top} \mathbf{H}\hat{\mathbf{A}}, \quad\text{where}\quad \mathbf{S}=\mathbf{R} + \hat{\mathbf{B}}^{\top} \mathbf{H}\hat{\mathbf{B}},
\end{equation}
so that the term $\hat{\mathbf{A}}^{\top} \mathbf{H} \hat{\mathbf{B}} (\mathbf{R}+ \hat{\mathbf{B}}^{\top} \mathbf{H} \hat{\mathbf{B}})^{-1} \hat{\mathbf{B}}^{\top} \mathbf{H} \hat{\mathbf{A}}$ in the modified DARE \eqref{eq:DAREbeta}
can be equivalently written as
\begin{equation} \label{eq:KSK}
\hat{\mathbf{A}}^{\top} \mathbf{H} \hat{\mathbf{B}} \mathbf{S}^{-1} \hat{\mathbf{B}}^{\top} \mathbf{H} \hat{\mathbf{A}} =  \hat{\mathbf{A}}^{\top}\mathbf{H}\hat{\mathbf{B}}\mathbf{K} = \mathbf{K}^{\top} \hat{\mathbf{B}}^{\top} \mathbf{H} \hat{\mathbf{A}} =\mathbf{K}^{\top} \mathbf{S} \mathbf{K},
\end{equation}
and the modified DARE \eqref{eq:DAREbeta} can be rewritten as
\begin{equation} \label{eq:DAREbetaB}
\hat{\mathbf{A}}^{\top} \mathbf{H} \hat{\mathbf{A}} - \beta^2\mathbf{H} - \mathbf{K}^{\top} \mathbf{S} \mathbf{K} + \mathbf{Q} = \mathbf{0}_{n \times n}.
\end{equation}
From this, we get the identity
\begin{equation} \label{eq:APAQKSK}
\hat{\mathbf{A}}^{\top} \mathbf{H} \hat{\mathbf{A}} - \beta^2\mathbf{H} = - \mathbf{Q} + \mathbf{K}^{\top} \mathbf{S}\mathbf{K}.
\end{equation}

Now, consider the Lyapunov function candidate $V(\mathbf{x})=\mathbf{x}^{\top} \mathbf{H} \mathbf{x}$.
We will prove that 
\begin{equation} \label{eq:Vkpu}
    V(\mathbf{x}_{k+1}) \le \beta^2 V(\mathbf{x}_k),  \quad k=0,1,\dots, 
\end{equation}
independently of the sequence $v_k$, provided that $v_k\in \mathcal{V}$, so that
\begin{equation} \label{eq:Vkbound}
    V(\mathbf{x}_{k}) \le \beta^{2k} V(\mathbf{x}_0),\quad k=0,1,\dots. 
\end{equation}
From this, recalling that $h_{\min}\Vert \mathbf{x}\Vert^2  \le \mathbf{x}^{\top} \mathbf{H} \mathbf{x} \le h_{\max}\Vert \mathbf{x}\Vert^2$, we easily get
\begin{equation}
    h_{\min}\Vert \mathbf{x}_{k}\Vert^2 \le \beta^{2k} h_{\max} \Vert \mathbf{x}_0\Vert^2,
\end{equation}
from which, taking the square roots of both terms of the inequality, the inequality \eqref{eq:expbound} follows.

Thus, to prove the Theorem, i.e., the inequalities \eqref{eq:expbound}, it is sufficient to prove the inequalities \eqref{eq:Vkpu}.
These can be rewritten as
\begin{equation}
    V(\mathbf{x}_{k+1}) - \beta^2 V(\mathbf{x}_k)\le 0, \quad k=0,1,\dots,
\end{equation}
or
\begin{equation}
    \mathbf{x}_{k+1}^{\top} \mathbf{H}\mathbf{x}_{k+1} - \beta^2 \mathbf{x}_{k}^{\top} \mathbf{H}\mathbf{x}_{k}\le 0, \quad k=0,1,\dots.
\end{equation}
Exploiting the system equation \eqref{eq:DLTSysClosed}, these inequalities change to the following
\begin{equation} \label{eq:seqineq}
\mathbf{x}^{\top}_{k} \big( (\hat{\mathbf{A}} - v_k \hat{\mathbf{B}}\mathbf{K})^{\top} \mathbf{H} (\hat{\mathbf{A}} - v_k \hat{\mathbf{B}}\mathbf{K}) - \beta^2 \mathbf{H}\big) \mathbf{x}_{k}\le 0, %\quad k=0,1,\dots.
\end{equation}
where $v_k\in[\underline{v},\overline{v}]$. 
It is clear that all inequalities \eqref{eq:seqineq} are verified if the following inequality is true:
\begin{equation} \label{eq:mainineq}
(\hat{\mathbf{A}} - v \hat{\mathbf{B}}\mathbf{K})^{\top} \mathbf{H} (\hat{\mathbf{A}} - v \hat{\mathbf{B}}\mathbf{K}) - \beta^2 \mathbf{H} \le 0, \quad \forall v\in[\underline{v},\overline{v}].
\end{equation}
Upon carrying out the multiplications in the left-hand side term of the inequality and considering 
the identities \eqref{eq:KSK} and \eqref{eq:APAQKSK}, we have
%\begin{equation} % \label{eq:mainineq}
% \begin{aligned}
% &(\mathbf{A} - v \mathbf{B}\mathbf{K})^{\top}  \mathbf{H} (\mathbf{A} - v \mathbf{B}\mathbf{K}) - \beta^2 \mathbf{H} \\
% & = \mathbf{A}^{\top}\mathbf{H}  +\mathbf{H}\mathbf{A} -  \beta^2 \mathbf{H} - v \mathbf{K}^{\top} \mathbf{B}^{\top} \mathbf{H} \mathbf{A} - v \mathbf{A}^{\top}\mathbf{H}\mathbf{B}\mathbf{K} +v^2 
% \mathbf{K}^{\top} \mathbf{B}^{\top} \mathbf{H}\mathbf{B}\mathbf{K} \\
% & = \mathbf{A}^{\top}\mathbf{H} +\mathbf{H}\mathbf{A} - \beta^2 \mathbf{H} - 2v \mathbf{K}^{\top} \mathbf{S} \mathbf{K} + v^2 \mathbf{K}^{\top} \mathbf{B}^{\top} \mathbf{H}\mathbf{B}\mathbf{K}    \\ 
% & = -\mathbf{Q} + \mathbf{K}^{\top} \mathbf{S} \mathbf{K} - 2v \mathbf{K}^{\top} \mathbf{S} \mathbf{K} + v^2 \mathbf{K}^{\top} \mathbf{B}^{\top} \mathbf{H}\mathbf{B}\mathbf{K}\\
% & = -\mathbf{Q} + \mathbf{K}^{\top} \big( (1- 2v) \mathbf{S} + v^2 \mathbf{B}^{\top} \mathbf{H}\mathbf{B}\big)\mathbf{K}.\\
% \end{aligned}
\begin{equation} % \label{eq:mainineq} 
\begin{aligned}
    &(\hat{\mathbf{A}} - v \hat{\mathbf{B}}\mathbf{K})^{\top}  \mathbf{H} (\hat{\mathbf{A}} - v \hat{\mathbf{B}}\mathbf{K}) - \beta^2 \mathbf{H} \\
    &= \hat{\mathbf{A}}^{\top}\mathbf{H} \hat{\mathbf{A}} -  \beta^2 \mathbf{H} 
     - v \mathbf{K}^{\top} \hat{\mathbf{B}}^{\top} \mathbf{H} \hat{\mathbf{A}} - v \hat{\mathbf{A}}^{\top} \mathbf{H} \hat{\mathbf{B}} \mathbf{K} \\
    &\quad + v^2 \mathbf{K}^{\top} \hat{\mathbf{B}}^{\top} \mathbf{H} \hat{\mathbf{B}} \mathbf{K} \\
    &= \hat{\mathbf{A}}^{\top}\mathbf{H} \hat{\mathbf{A}} - \beta^2 \mathbf{H} 
     - 2v \mathbf{K}^{\top} \mathbf{S} \mathbf{K} + v^2 \mathbf{K}^{\top} \hat{\mathbf{B}}^{\top} \mathbf{H} \hat{\mathbf{B}} \mathbf{K} \\ 
    &= -\mathbf{Q} + \mathbf{K}^{\top} \mathbf{S} \mathbf{K}
     - 2v \mathbf{K}^{\top} \mathbf{S} \mathbf{K} + v^2 \mathbf{K}^{\top} \hat{\mathbf{B}}^{\top} \mathbf{H} \hat{\mathbf{B}} \mathbf{K} \\
    &= -\mathbf{Q} + \mathbf{K}^{\top} \big( (1- 2v) \mathbf{S} + v^2 \hat{\mathbf{B}}^{\top} \mathbf{H} \hat{\mathbf{B}} \big) \mathbf{K}.
\end{aligned}
\end{equation}

%\end{equation}
Recalling that $\mathbf{S}=\mathbf{R} + \hat{\mathbf{B}}^{\top} \mathbf{H}\hat{\mathbf{B}}$, we get % the equivalent forms for the left-hand-side of 
\eqref{eq:mainineq}:
\begin{equation} \label{eq:equiQK}
\begin{aligned} 
(\hat{\mathbf{A}} - &v \hat{\mathbf{B}}\mathbf{K})^{\top}  \mathbf{H} (\mathbf{A} - v \hat{\mathbf{B}}\mathbf{K}) - \beta^2 \mathbf{H} \\
 & = -\mathbf{Q}+\mathbf{K}^{\top}\big( (v-1)^2 \hat{\mathbf{B}}^{\top} \mathbf{H}\hat{\mathbf{B}} + (1-2v)\mathbf{R}\big)\mathbf{K}. 
 \end{aligned}
\end{equation}
Thus, the inequality \eqref{eq:mainineq} can be  written equivalently in the form \eqref{eq:ineqQKK} of Lemma \ref{Lem:QKGKpos}, from which we know that
there exists a nonempty interval $[\underline{v},\overline{v}]$ such that the matrix \eqref{eq:equiQK} is negative semidefinite.
It follows that all inequalities \eqref{eq:seqineq} are satisfied as long as $v_k\in[\underline{v},\overline{v}]$, and this implies inequality \eqref{eq:expbound}, and the Theorem is proved.
\end{proof}

\begin{remark}
    As discussed in Theorem \ref{Thm:mainTheo}, $\hat{\mathbf{B}}$ is essential for verifying \eqref{eq:ineqQKK} and determining $v \in [\underline{v}, \overline{v}]$ to ensure stability. If $\hat{\mathbf{B}}$ is unknown, it can be estimated from collected data using  
    \begin{equation}
        \begin{bmatrix} \hat{\mathbf{B}} & \hat{\mathbf{A}} \end{bmatrix} = \mathbf{X}_1 \begin{bmatrix}
            \mathbf{U}_0 \\
            \mathbf{X}_0
        \end{bmatrix}^{\dagger}.
    \end{equation}
    Since any interval $(v_m, v_M) \subseteq [\underline{v}, \overline{v}]$ with $v_m < 1 < v_M$ can replace $\mathcal{V}$ in Algorithm~\ref{algo:state perturbationsfreeDeePO}, choosing a narrower interval is preferable, as it reduces control deviations from the optimum and improves stability margins. \hfill$\square$
\end{remark}

\section{Simulations}
\label{sec:Simulations}
To evaluate Algorithm~\ref{algo:state perturbationsfreeDeePO}, the open loop and controllable LTI system in~\cite{zhao2024data} is utilized in numerical simulations, with
\begin{equation}
\label{eq:simusys}
A =
\begin{bmatrix}
-0.13 & 0.14 & -0.29 & 0.28 \\
0.48 & 0.09 & 0.41 & 0.30 \\
-0.01 & 0.04 & 0.17 & 0.43 \\
0.14 & 0.31 & -0.29 & -0.10
\end{bmatrix}
\!\!,\quad \hspace{-0.2cm}
B =
\begin{bmatrix}
1.63 & 0.93 \\
0.26 & 1.79 \\
1.46 & 1.18 \\
0.77 & 0.11
\end{bmatrix}.
\end{equation}
First, a persistently exciting input, $\mathbf{U}_0$, is used to generate the offline state data $\mathbf{X}_0$ and $\mathbf{X}_1$ from the system dynamics in \eqref{eq:system}. The noise is sampled from a normal distribution, $\boldsymbol{\omega}_k=\mathcal{N}(0,\sigma_{\boldsymbol{\omega}})$ and the offline input is sampled from  $\mathbf{u}_k=\mathcal{N}(0,\sigma_u)$ with $\sigma_{\boldsymbol{\omega}}=\sigma_{\mathbf{u}}=0.01$. 
% \vitt{I'm puzzled by the choice of such a 'small' (in magnitude) input sequence to generate data. Is this adopted to let Algo 2 work with nice numbers?}
The offline data consists of 8 time samples. Since we consider an open-loop stable system, the initial policy is kept at zero, i.e, $\mathbf{K}_t = \mathbf{0}_{m \times n}$. The parameters of PFDeePO are chosen as $\gamma=0.1$, $\delta=0.1$,$\eta=10^{-4}$, and the interval of $\nu$ is chosen as $\underline{v}=0.5$ and $\overline{v}=1.5$, which satisfies \eqref{eq:ineqQKK}. DeePO, introduced in Algorithm~\ref{Algo:DeePO}, is used as a benchmark, with $\eta = 10^{-4}$ and a probing noise signal $\mathbf{e}$ sampled from a zero mean normal distribution with $\sigma_\mathbf{e}=0.1$. At sample $k=15$, a disturbance is induced in the states using a uniform random value. 

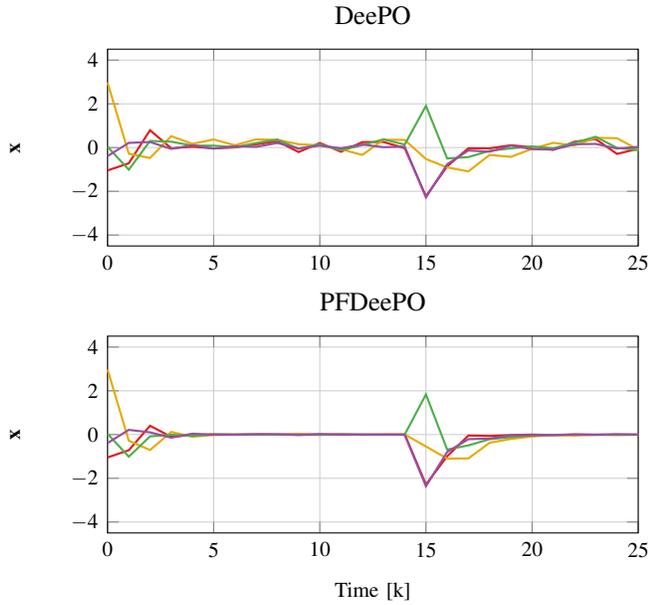
\begin{figure}
    \centering
    \input{Figures/stateEval}
    \caption{Evolution of the states $\mathbf{x}$, using DeePO (top) and PFDeePO (bottom).}
    \label{fig:stateEvol}
\end{figure}

In Fig.~\ref{fig:stateEvol}, we present the results of running both algorithms on the system. As observed, once the states reach equilibrium, PFDeePO does not introduce further perturbations. In contrast, the probing noise in DeePO continuously disturbs the states, inducing oscillations that increase control effort.

% \vitt{In the text and directly in Fig. 1 (title or caption) specify that Algo 1 is the existent DeePO and Algo 2 is the novel PFDeePO (or any other acronym you opted for).}

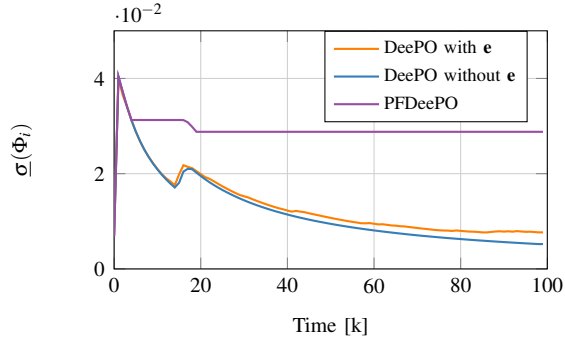
\begin{figure}
    \centering
    \input{Figures/minsvd}
    \caption{$\underline{\sigma}(\Phi_i)$ for PFDeePO and DeePO, with and without probing noise.}
    \label{fig:minsvd}
\end{figure}

Fig.~\ref{fig:minsvd} illustrates evolution of $\underline{\sigma}(\Phi_i)$. It is evident that in DeePO, probing noise is essential to maintaining the full rank of $\Phi$, as its minimum singular value otherwise approaches zero. However, with PFDeePO, $\underline{\sigma}(\Phi_i)$ remains consistently above zero throughout the control period, ensuring the rank condition is met without perturbing the system states.

%online_noise = randn(n,num)
%offline_noise = randn(n, T)
%probing_noise = randn(m,num); when comparing
%offline_input = randn(m,T);

%\addtolength{\textheight}{-16.5cm} 

\section{Conclusion}
\label{sec:Conclusion}
Data-driven control design has gained significant attention, but the field 
remains in its early stages, with several open challenges. Among recent 
approaches for LTI systems, DeePO is a promising method that refines control 
gains online through adaptive updates. However, its reliance on probing noise 
to ensure persistency of excitation can degrade performance and induce state 
perturbations.
This article introduces PFDeePO, which alleviates these issues. 
Simulations show that PFDeePO outperforms conventional DeePO in terms of 
control performance.
Future work includes extending PFDeePO to LTV and nonlinear systems and 
optimizing its design parameter selection.

% \bigskip
% \costanzo{ Alternative proof of Lemma \ref{lemma:rankphi}. }

% Write $\Phi$ defined in \eqref{eq:covariance_parametrization} only as a function of $t_2$, since $t_1$ is fixed:
% \begin{equation}
% \Phi(t_2) = \frac{1}{t_1 + t_2} 
% (\mathcal{D}_{1}\mathcal{D}_{1}^{\top} + \mathcal{D}_{2}\mathcal{D}_{2}^{\top}).
% \end{equation}
% Since $\mathcal{D}_{2}$ is not full row rank for all $t_2$, there exists a vector $\mathbf{a}\in \mathbb{R}^{n+m}$ such that $\mathbf{a}^\top \mathcal{D}_{2}=0$.
% Thus
% \begin{equation}
% \begin{aligned}    
%     \mathbf{a}^\top \Phi(t_2)\mathbf{a} & = 
%     \frac{1}{t_1 + t_2} (\mathbf{a}^\top \mathcal{D}_{1}\mathcal{D}_{1}^{\top} \mathbf{a}
%     +\mathbf{a}^\top \mathcal{D}_{2}\mathcal{D}_{2}^{\top} \mathbf{a})      \\
%     & =   \frac{1}{t_1 + t_2} \mathbf{a}^\top \mathcal{D}_{1}\mathcal{D}_{1}^{\top} \mathbf{a}.
% \end{aligned}
% \end{equation}
% From this, it is clear that 
% \begin{equation} \label{eq:limitaPhia}
%     \lim_{t_2\to\infty }\mathbf{a}^\top \Phi(t_2)\mathbf{a} = 0.
% \end{equation}
% Since the smallest singular value of $\Phi(t_2)$ is such that
% \begin{equation}
% \underline{\sigma}(\Phi(t_2)) \Vert \mathbf{a} \Vert^2 \le \mathbf{a}^\top \Phi(t_2)\mathbf{a},\quad \forall t_2\in\mathbb{N},
% \end{equation}
% then the limit \eqref{eq:limitaPhia} implies
% \[
% \lim_{t_2 \to \infty} \underline{\sigma}\big(\Phi(t_2)\big) \to 0,
% \]
% that is the thesis.
% \costanzo{ End of the proof of Lemma \ref{lemma:rankphi}. }

\bibliographystyle{IEEEtran}
\bibliography{biblio}

  % This command serves to balance the column lengths
                                  % on the last page of the document manually. It shortens
                                  % the textheight of the last page by a suitable amount.
                                  % This command does not take effect until the next page
                                  % so it should come on the page before the last. Make
                                  % sure that you do not shorten the textheight too much.

%%%%%%%%%%%%%%%%%%%%%%%%%%%%%%%%%%%%%%%%%%%%%%%%%%%%%%%%%%%%%%%%%%%%%%%%%%%%%%%%

\end{document}

%% file: Figures/stateEval.tex
\definecolor{c1}{RGB}{228,26,28}
\definecolor{c2}{RGB}{55,126,184}
\definecolor{c3}{RGB}{77,175,74}
\definecolor{c4}{RGB}{152,78,163}
\definecolor{c5}{RGB}{255,127,0}
\definecolor{c6}{RGB}{230,171,2}

%\definecolor{c1}{RGB}{27,158,119}
%\definecolor{c2}{RGB}{217,95,2}
%\definecolor{c3}{RGB}{117,112,179}
%\definecolor{c4}{RGB}{231,41,138}
%\definecolor{c5}{RGB}{102,166,30}
%\definecolor{c6}{RGB}{230,171,2}

\begin{tikzpicture}
\begin{groupplot}[
group style={group size=1 by 2,
    ylabels at=edge left,
    yticklabels at=edge left,
    xlabels at=edge bottom,
    vertical sep=1.2cm
},
width=\columnwidth,
height=4.2cm,
grid = both,
grid style={line width=.1pt, draw=gray!20},
major grid style={line width=.2pt,draw=gray!40},
ylabel style={align=center, font=\footnotesize},
xlabel style={font=\footnotesize},
tick label style={font=\footnotesize},
xlabel = {Time [k]},
enlargelimits=false,
xmin=0, xmax=25,
]
\nextgroupplot[
title= DeePO,
ylabel={$\mathbf{x}$}, 
ymin=-4.5,
ymax=4.5,
ytick={-4,-2,...,4}
]
\addplot[c1,thick] table [x=Time, y=x1, col sep=comma] {./Figures/data/alg1.csv};
\addplot[c6,thick] table [x=Time, y=x2, col sep=comma] {./Figures/data/alg1.csv};
\addplot[c3,thick] table [x=Time, y=x3, col sep=comma] {./Figures/data/alg1.csv};
\addplot[c4,thick] table [x=Time, y=x4, col sep=comma] {./Figures/data/alg1.csv};

\nextgroupplot[
title= PFDeePO,
ylabel={$\mathbf{x}$}, 
ymin=-4.5,
ymax=4.5,
ytick={-4,-2,...,4}
]
\addplot[c1,thick] table [x=Time, y=x1, col sep=comma] {./Figures/data/alg2.csv};
\addplot[c6,thick] table [x=Time, y=x2, col sep=comma] {./Figures/data/alg2.csv};
\addplot[c3,thick] table [x=Time, y=x3, col sep=comma] {./Figures/data/alg2.csv};
\addplot[c4,thick] table [x=Time, y=x4, col sep=comma] {./Figures/data/alg2.csv};
\end{groupplot}
\end{tikzpicture}

%% file: Figures/minsvd.tex
\definecolor{c1}{RGB}{228,26,28}
\definecolor{c2}{RGB}{55,126,184}
\definecolor{c3}{RGB}{77,175,74}
\definecolor{c4}{RGB}{152,78,163}
\definecolor{c5}{RGB}{255,127,0}
\definecolor{c6}{RGB}{230,171,2}

\begin{tikzpicture}
\begin{axis}[
width=0.85\columnwidth,
height=0.55\columnwidth,
%title = Minimum singular value of $\Phi$, 
grid = both,
grid style={line width=.1pt, draw=gray!20},
major grid style={line width=.2pt,draw=gray!40},
xlabel style={font = \footnotesize},
tick label style={font=\footnotesize},
xlabel = {Time [k]},
xmin=0, xmax=100,
xtick={0,20,...,100},
ylabel={$\underline{\sigma}(\Phi_i)$},
ylabel style={align=center, font=\footnotesize},
ylabel shift = {1000pt},
ymin=0,
ymax=0.05,
legend cell align={left},
legend entries={
DeePO with $\mathbf{e}$,
DeePO without $\mathbf{e}$,
PFDeePO},
legend style={
    at={(0.72, 1)}, 
    anchor=north, 
    font=\scriptsize, 
    legend columns=1,
    row sep=1.2pt, % Adjusts the vertical spacing between rows
    nodes={inner sep=1.5pt}
}
]
\addplot[c5, thick] table [x=Time, y=minsvd, col sep=comma] {Figures/data/alg1.csv};
\addplot[c2, thick] table [x=Time, y=minsvd, col sep=comma] {Figures/data/alg1NoProbe.csv};
\addplot[c4, thick] table [x=Time, y=minsvd, col sep=comma] {Figures/data/alg2.csv};
\end{axis}
\end{tikzpicture}